\documentstyle[aps,twocolumn,prl,epsf]{revtex}

\newcommand{\PRB}{Phys.\ Rev.\ B}

\newcommand{\PRL}{Phys.\ Rev.\ Lett.}

\begin{document}
\draft

\title{Numerical Renormalization Group Study
       of the O(3)-symmetric Anderson Model}

\author{R.\ Bulla \and A.\ C.\ Hewson} 
\address{Department of Mathematics, Imperial College, 180 Queen's Gate,
London SW7 2BZ, U. K.}
\date{\today}
\maketitle

\begin{abstract}
We use the numerical renormalization group method to study the O(3)-symmetric
version of the impurity Anderson model of  Coleman and Schofield.
This model is of general interest because  it
displays both  Fermi liquid and non-Fermi liquid behaviour, and in the
large $U$
limit can be related to the compactified two channel Kondo model of Coleman,
Ioffe and Tsvelik.
We calculate the thermodynamics for a parameter range
which covers the full range of behaviour of the model. 
We find a non-Fermi liquid fixed point in the isotropic case which is unstable
with respect to channel anisotropy.


\end{abstract}

\section{Introduction}

The unusual experimental results observed in many aspects of the behaviour of
the high temperature superconductors
and some heavy fermion
systems \cite{NFL}, has led to the conjecture that this behaviour 
can best be explained by 
some form of non-Fermi liquid theory. For strong correlation models that 
have been put forward to describe these systems it is very difficult to test the 
stability of the Fermi liquid state due to  the strong local interaction $U$.
Perturbative methods are unreliable for large values of $U$, and non-perturbative 
techniques are in general difficult to apply and subject to uncertainties in the
approximations that have to be used. With direct numerical methods such as
Monte Carlo calculations there are very severe size limitations (and also the
`sign problem') that make it difficult to obtain results for excitations for the
thermal energy scale that are required  to examine these issues.
It is
of some interest, therefore, to study strong correlation  impurity models that 
display similar types of  non-Fermi liquid behaviour as accurate methods
for predicting
the behaviour of these systems have been developed, including many exact
solutions.

The non-Fermi liquid behaviour of the  impurity  two channel Kondo model
has already been studied
extensively and has been used as a basis for explaining  the behaviour of
certain heavy fermion compounds \cite{Cox95}. 
It has also been used to interpret the  experimental results  for certain
two level tunnelling systems \cite{Ral92}. 

Recently a modified form of the Anderson impurity
model has been formulated by Coleman and Schofield \cite{Col95a} which 
also displays
non-Fermi liquid behaviour. In the large $U$ limit the model  can be related
to a form of the two channel Kondo model \cite{Col95a}, and for small $U$ 
there is a parameter regime where
it displays marginal Fermi liquid behaviour \cite{Col95,Zha96}.
This model is simple enough so that its behaviour can be studied using the
numerical renormalization group techniques (NRG) 
that were used to obtain definitive results for the
standard Anderson model \cite{Kri80}. This modified Anderson 
model, the O(3)-symmetric Anderson model (O(3)-AM), has the form,
\begin{eqnarray}
  H_{\rm O(3)-AM} &=& \sum_\sigma \varepsilon_{\rm f} f^\dagger_\sigma
                             f_\sigma
                 + U  f^\dagger_\uparrow f_\uparrow
                       f^\dagger_\downarrow f_\downarrow \nonumber\\
         &+&               
            \sum_{ \sigma}\int_{-1}^1 {\rm d} \varepsilon \, \varepsilon
                    a^\dagger_{\varepsilon \sigma} a_{\varepsilon \sigma} \nonumber \\
           &+&  \sum_{ \sigma} \int_{-1}^1 {\rm d} \varepsilon \,
                   V(\varepsilon) \sqrt{\rho(\varepsilon)}
                    \Big( f^\dagger_\sigma
                   a_{\varepsilon \sigma}  +
                     a^\dagger_{\varepsilon \sigma} f_\sigma \Big)  \nonumber \\
      -  \int_{-1}^1 {\rm d} \varepsilon 
      & & \!\!\!\!\!\!\!\! V_{\rm a}(\varepsilon) \sqrt{\rho(\varepsilon)}
         \left( a_{\varepsilon\downarrow}^{\dagger}f_{\downarrow}
      +  a_{\varepsilon\downarrow}f_{\downarrow} +
         f_{\downarrow}^\dagger a_{\varepsilon\downarrow}
      + f_{\downarrow}^{\dagger} a_{\varepsilon\downarrow}^\dagger \right)\ .
\label{eq:CAM}
\end{eqnarray}
Apart from the term with the anomalous hybridization matrix element $V_{\rm a}$ 
this is the standard
Anderson model for a localized f level hybridized with a conduction band and an 
on-site interaction $U$.
The model is taken to be particle-hole symmetric with $\varepsilon_{\rm f}=-U/2$
and $\varepsilon_k = -\varepsilon_{-k}$.
Due to the antisymmetry of the dispersion in $H_{\rm O(3)-AM}$, the
hybridization $V(\varepsilon)$ and the anomalous hybridization
$V_{\rm a}(\varepsilon)$ have to be antisymmetric (and are here
chosen to be constant).
\begin{equation}
   V(\varepsilon) = \left\{
            \begin{array}{rcl}
               V & : & \varepsilon > 0 \\
               -V & : & \varepsilon < 0
             \end{array}   \right. \ \
   V_{\rm a}(\varepsilon) = \left\{
            \begin{array}{rcl}
               V_{\rm a} & : & \varepsilon > 0 \\
               -V_{\rm a} & : & \varepsilon < 0
             \end{array}   \right. \ .
\end{equation} 
The extra anomalous hybridization term
is one in which neither particles nor spin is conserved.
It reduces the symmetry of the standard Anderson model at 
particle-hole symmetry from O(4) to O(3) (the O(3)-symmetry is best seen in
the Majorana Fermion formulation \cite{Col95,Zha96}).

The motivation for its introduction is that in the large $U$ limit the model can
be transformed by a
Schrieffer-Wolff transformation \cite{Col95a} into a localized s-d type of model of the form
\begin{equation}
   H_{\rm CKM} = \sum_{k\sigma} \varepsilon_k c^\dagger_{k\sigma}
             c_{k\sigma} + \vec{S}\cdot \left[ J_1 \vec{s}(0) + J_2 \vec{\tau}
                (0) \right] \ .
\end{equation}
 The  operators $\vec{s}(0)$ are the  spin operators for the conduction electron
 states
defined in terms of the conduction electron creation and annihilation operators 
($c^\dagger_{k\sigma}$, $c_{k\sigma}$) in the
usual way,
$$
   s^+ = c^\dagger_\uparrow c_\downarrow, \quad
   s^- = c^\dagger_\downarrow c_\uparrow, $$
\begin{equation}   s_z = \frac{1}{2}\left( c^\dagger_\uparrow c_\uparrow
              - c^\dagger_\downarrow c_\downarrow  \right).
\end{equation}

 The  $\vec{\tau}(0)$ operators are  isospin operators for the conduction 
electrons and these  are
defined by
$$
   \tau^+ = c^\dagger_\uparrow c^\dagger_\downarrow, \quad
   \tau^- = c_\downarrow c_\uparrow ,$$
 \begin{equation}  \tau_z = \frac{1}{2}\left( c^\dagger_\uparrow c_\uparrow
              + c^\dagger_\downarrow c_\downarrow -1 \right) \ .
\end{equation}
Both sets of operators satisfy the usual SU(2) commutation relations.
The spin and isospin of the conduction electrons are coupled to the impurity
 spin $\vec{S}$ individually with
coupling constants $J_1$ and $J_2$, respectively. From the Schrieffer-Wolff
 transformation \cite{Col95a} these coupling constants are given by 
\begin{equation}
    J_1 = \frac{4V(V-V_{\rm a})}{U},\quad
    J_2 = \frac{4VV_{\rm a}}{U} \ .
\end{equation}
This localized  model eq.\ (3), which was introduced by Coleman, Ioffe and Tsvelik
 \cite{Col95}, has a
 form very similar to the two
channel Kondo model and it has been referred to as the 
compactified two channel
Kondo model or ($\tau$-$\sigma$)-model. Though it differs from the two channel Kondo model in that the
 impurity spin cannot be
overscreened by the conduction electrons (the states with spin and the states 
with isopin are mutually
exclusive and so the cannot both be used to screen the impurity at the same time)
it has been argued that
it has the same low energy fixed point as the two channel Kondo model \cite{Col95}.
The channel isotropic case ($J_1=J_2$ in eq.\ (6)), where we expect
non-Fermi liquid behaviour, corresponds to $V_{\rm a}=V/2$.

It is
clear that the general O(3)-AM has a rich
range of behaviour and may provide theoretical insights into the controversial
questions as the non-Fermi liquid behaviour observed in some strongly
correlated systems. Due to the particle non-conserving terms it is unlikely
that the Bethe ansatz technique, which has been used to solve many of the
strong correlation impurity models,
can provide a solution for this case. 

The numerical renormalization group method,
however, as originally developed
by Wilson for the Kondo problem \cite{Wil75}, 
and further used by Krishnamurthy et al.\ \cite{Kri80} to obtain definite results 
for the Anderson model in all parameter regimes can be used.
In this paper we apply this approach and give details of the calculation and
results for the
thermodynamic behaviour (a brief summary of this work is given in
\cite{Bul97}). 
The only
slight complication in applying the NRG to the O(3)-AM 
is that the many-body states obtained in the iterative
diagonalization cannot be classified in terms of  quantum numbers associated with
the particle number and spin individually. There is a total spin operator $\vec{
T}$, which is the sum of the spin and isospin, which does commute with the
Hamiltonian. It is defined by
\begin{equation}
T_z=\sum_n(s_{n,z}+\tau_{n,z})\quad , \  T^{\pm}=\sum_n(s^{\pm}_{n}+\tau^{\pm}_{n}),
\end{equation}
where the sum runs over the impurity and conduction electron states. For
 perturbational calculations it is
useful to express the Hamiltonian in terms of Majorana fermion operators
 \cite{Col95a,Col95}.  In the Majorana fermion form  it is clear
that the non-Fermi liquid situations which develop for $V_{\rm a}=V/2$
 are due to the fact that one of the
Majorana fermions associated with the impurity is unhybridized. This gives a
 local zero mode and the scattering
with the conduction electrons via the $U$ term leads to singularities and the
 breakdown of Fermi liquid theory.
There is no advantage, however, in the numerical renormalization group approach 
in using the Majorana fermion
representation so we retain the model in the form eq.\ (\ref{eq:CAM}). 
We do need, however,
 to re-express the Hamiltonian along standard lines for the numerical
renormalization group approach and introduce a discrete basis for the conduction
 electrons for the iterative sequence of diagonalizations that is used.

\section{Details of the Numerical Renormalization Group Analysis}

In order to apply the NRG method on the O(3)-AM eq.\ (\ref{eq:CAM}),
its Hamiltonian has to be cast in a semi-infinite chain form.
This mapping is described in detail in \cite{Kri80}.
The logarithmic discretization (with the discretization parameter 
$\Lambda\! >\! 1$), the definition of the discrete operators
$a_{np\sigma}$ and  $b_{np\sigma}$
and the transformation of the free conduction electron term
are identical to the treatment of the standard single impurity
Anderson model (SIAM) \cite{Kri80}.
The  $p\ne 0$ fermions again do not couple directly to the impurity and are neglected.

Due to the antisymmetry of $V(\varepsilon)$, there is a minus sign
in the definition of the conduction electron operator at the
impurity site
\begin{equation}
  g_{0\sigma} =  \frac{\sqrt{1-\Lambda^{-1}}}{\sqrt{2}}
     \sum_n \Lambda^{-n/2} \left(a_{n\sigma} - b_{n\sigma}\right) \ \ .
\end{equation}
In contrast to the combination $a_{n\sigma} + b_{n\sigma}$
one obtains in the standard case.

The discretized version of $H_{\rm O(3)-AM}$ then reads
\begin{eqnarray}
H_{\rm O(3)-AM} &=& \sum_\sigma \varepsilon_{\rm f} f^\dagger_\sigma
                             f_\sigma
                 + U  f^\dagger_\uparrow f_\uparrow
                       f^\dagger_\downarrow f_\downarrow
                          \nonumber \\
            &+& \frac{1}{2} (1 + \Lambda^{-1} )
                \sum_{n\sigma} \Lambda^{-n} \left(
                  a^\dagger_{n\sigma} a_{n\sigma} -
                  b^\dagger_{n\sigma} b_{n\sigma} \right)  \nonumber \\
            &+& V\sum_\sigma \left( f^\dagger_{\sigma} g_{0\sigma} +
                  g^\dagger_{0\sigma} f_{\sigma} \right) \\
   &-&  V_{\rm a} \left( g_{0\downarrow}^{\dagger}f_{\downarrow}
      +  g_{0\downarrow}f_{\downarrow} + f_{\downarrow}^\dagger g_{0\downarrow}
      + f_{\downarrow}^{\dagger}g_{0\downarrow}^\dagger \right) 
 \ \ .\nonumber
\end{eqnarray}
The minus sign in (8) does not affect the semi-infinite chain form of
the conduction electron part
\begin{equation}
   H_{\rm c} = \sum_{\sigma n=0}^{\infty} t_n \left(
       g^\dagger_{n\sigma} g_{n+1\sigma} +
                  g^\dagger_{n+1\sigma} g_{n\sigma} \right) \ \ ,
\end{equation}
with
\begin{eqnarray}
   t_n & &= \frac{1}{2} \left( 1 + \Lambda^{-1} \right)
                 \left( 1 - \Lambda^{-n-1} \right) \nonumber \\ 
       &\times&           \left( 1 - \Lambda^{-2n-1} \right)^{-1/2}
                 \left( 1 - \Lambda^{-2n-3} \right)^{-1/2}
                \Lambda^{-n/2} \ \ .
\end{eqnarray}

As in the standard case \cite{Kri80} we diagonalize iteratively a sequence
of Hamiltonians $H_N$ with
\begin{equation}
      H_{-1} = \frac{1}{\Lambda} \left[
            \sum_\sigma \varepsilon_{\rm f} f^\dagger_\sigma
                             f_\sigma
                 + U  f^\dagger_\uparrow f_\uparrow
                       f^\dagger_\downarrow f_\downarrow
      \right] ,
\end{equation}
\begin{equation}
      H_{N+1} = \sqrt{\Lambda} H_N + \Lambda^{N/2} \sum_\sigma
              t_N
     \Big(
                g^\dagger_{N \sigma} g_{N+1 \sigma}
             +   g^\dagger_{N+1 \sigma} g_{N\sigma}  \Big) ,
\end{equation}
so that
\begin{equation}
      H_{\rm O(3)-AM} = \lim_{N\to\infty} \Lambda^{-(N-1)/2} H_N \ \ \ .
\end{equation}

At this point it is convenient to generalize the Hamiltonian (9) by an
anomalous hybridization between {\it all} sites of the
semi-infinite chain.
Together with the $V_{\rm a}$-term we have
\begin{equation}
    H_{\rm a} = \sum_{n=-1}^\infty H_{\rm a,n}  \  ,
\end{equation}
\begin{eqnarray}
    H_{\rm a,n} &=& -t_{{\rm a},n}
      \left[ g_{n+1\downarrow}^{\dagger}g_{n\downarrow}
      +(-1)^{n+1}  g_{n+1\downarrow}g_{n\downarrow} 
     \right. \nonumber \\
     &+& \left. g_{n\downarrow}^\dagger g_{n+1\downarrow}
      + (-1)^{n+1} g_{n\downarrow}^{\dagger}g_{n+1\downarrow}^\dagger \right] \ \ \ \
\ ,
\end{eqnarray}
with $t_{{\rm a},-1} \equiv V_{\rm a}$ .
The $(-1)^{n+1}$ factors are due to the requirement that $T^2$
commutes with $H$. Obviously, all $t_{{\rm a},n}$ are zero for
$n\ge 0$, but the advantage of this generalization is that we can start the
iterative procedure by adding site $0$ to the impurity. Otherwise the
two-site problem would have to be the starting point.

Both particle number $Q$ and spin $\vec{S}$ are no more conserved quantities of the 
Hamiltonian (9) as soon as $V_{\rm a} \ne 0$. The only conserved quantity is
the total `spin plus isospin'
\begin{equation}
   \vec{T} = \vec{S} + \vec{\tau} \ \ \ \  \ \ ,
\end{equation}
with
\begin{eqnarray}
     \vec{\tau} &=& \vec{\tau}_{\rm f} + \sum_{n=0}^\infty \vec{\tau}_n \ \ \ \ \ ,\\
     \vec{S} &=& \vec{s}_{\rm f} + \sum_{n=0}^\infty \vec{s}_n 
\end{eqnarray}
and
\begin{eqnarray}
   \tau_n^+ &=& (-1)^{n+1} g^\dagger_{n\uparrow} g^\dagger_{n\downarrow} \ \ \ \ \ ,\nonumber \\
   \tau_n^- &=& (-1)^{n+1} g_{n\downarrow} g_{n\uparrow} \ \ \ \ \ ,\\
   \tau_{nz} &=& \frac{1}{2}\left( g^\dagger_{n\uparrow} g_{n\uparrow}
              + g^\dagger_{n\downarrow} g_{n\downarrow} -1 \right) 
                      \ \ \ \ \ .  \nonumber
\end{eqnarray}
The $n$-dependent definition of $\vec{\tau}_n$ is necessary to fulfill the
condition $[T^{+/-}, H ]=0$.

The starting point for the iterative diagonalization is the uncoupled impurity
(labeled by $-1$).
Writing the states in the form $\left\vert T,T_z,r \right>$
we get
\begin{eqnarray}
   \left\vert \frac{1}{2}, \frac{1}{2}, 1 \right>_{-1} &=& 
                \left\vert \uparrow \downarrow \right> \ \  ,\nonumber\\
   \left\vert \frac{1}{2}, -\frac{1}{2}, 1 \right>_{-1} &=& 
                  \left\vert 0 \right>\ \  ,\ \nonumber\\
   \left\vert \frac{1}{2}, \frac{1}{2}, 2 \right>_{-1} &=& 
                 \left\vert \uparrow  \right> \ \  ,\\
   \left\vert \frac{1}{2}, -\frac{1}{2}, 2 \right>_{-1} &=& 
                \left\vert  \downarrow \right> \ \  .\nonumber
\end{eqnarray}

Supposed that, at the $N$th step of the iteration procedure, the
Schr\"odinger equation
\begin{equation}
   H_N \left\vert T,T_z,r \right>_N = E_N (T,r) \left\vert T,T_z,r \right>_N
\end{equation}
has been solved, we can set up the Hamiltonian matrix for the
$(N+1)$st step.
First we define a basis for the enlarged system
\begin{equation}
  \begin{array}{rcl}
   \left\vert T,T_z,r;0 \right> &=& \left\vert T,T_z,r \right>_N \ \ ,\\ \vspace*{0cm}\\
   \left\vert T,T_z,r;\uparrow \right> &=&
             g^\dagger_{N+1\uparrow }\left\vert T,T_z,r \right>_N \ \ , \\ \vspace*{0cm}\\
   \left\vert T,T_z,r;\downarrow \right> &=&
             g^\dagger_{N+1\downarrow}\left\vert T,T_z,r \right>_N \ \ ,\\ \vspace*{0cm}\\
   \left\vert T,T_z,r;\uparrow \downarrow \right> &=&
      g^\dagger_{N+1\uparrow} g^\dagger_{N+1\downarrow}
      \left\vert T,T_z,r \right>_N  \ \ .
   \end{array}
\end{equation}
These states can be combined to give eigenstates of $T^2$ and $T_z$
\begin{equation}
  \begin{array}{rcl}
   \left\vert T,T_z,r;1 \right>_{N+1} &=&
              (-1)^N\sqrt{\frac{T+T_z}{2T}} \left\vert T-\frac{1}{2},
                      T_z-\frac{1}{2},r;\uparrow\downarrow \right> \\
                      & & \!\!\!\!\!+
              \sqrt{\frac{T-T_z}{2T}} \left\vert T-\frac{1}{2},
                      T_z+\frac{1}{2},r;0\right> \ \ ,\\ \\
   \left\vert T,T_z,r;2 \right>_{N+1} &=&
              \sqrt{\frac{T+T_z}{2T}} \left\vert T-\frac{1}{2},
                      T_z-\frac{1}{2},r;\uparrow \right> \\
                      & & \!\!\!\!\! +
              \sqrt{\frac{T-T_z}{2T}} \left\vert T-\frac{1}{2},
                      T_z+\frac{1}{2},r;\downarrow\right> \ \ ,\\ \\
   \left\vert T,T_z,r;3 \right>_{N+1} &=&
              \sqrt{\frac{T-T_z+1}{2T+2}} \left\vert T+\frac{1}{2},
                      T_z-\frac{1}{2},r;\uparrow \right> \\
                      & & \!\!\!\!\!-
              \sqrt{\frac{T+T_z+1}{2T+2}} \left\vert T+\frac{1}{2},
                      T_z+\frac{1}{2},r;\downarrow\right> \ \ ,\\ \\
   \left\vert T,T_z,r;4 \right>_{N+1} &=&
              \sqrt{\frac{T-T_z+1}{2T+2}} \left\vert T+\frac{1}{2},
                      T_z-\frac{1}{2},r;\uparrow\downarrow \right> \\
                      & & \!\!\!\!\!\!\!\!\!\!\!\!\!\!\!-
              (-1)^N\sqrt{\frac{T+T_z+1}{2T+2}} \left\vert T+\frac{1}{2},
                      T_z+\frac{1}{2},r;0\right> \ \ \ .\\
  \end{array}
\end{equation}

The Hamiltonian matrix
\begin{equation}
     H_T(ri,r^\prime j) =  \phantom{>}_{N+1}
     \left< T,T_z,r;i \right\vert H_{N+1}
        \left\vert T,T_z,r^\prime;j \right>_{N+1}
\end{equation}
now consists of three parts:
\begin{eqnarray}
     H_T(ri,r^\prime j)
    = \sqrt{\Lambda}  \phantom{>}_{N+1} \left< T,T_z,r;i \right\vert H_N
     \left\vert T,T_z,r^\prime;j \right>_{N+1} \nonumber \\
    + \ \Lambda^{N/2}\  \sum_\sigma
              t_N
    \phantom{>}_{N+1} \left<T,T_z,r;i \right\vert X
              \left\vert T,T_z,r^\prime;j \right>_{N+1} \nonumber \\
    +  \ \Lambda^{N/2} \ 
    \phantom{>}_{N+1} \left<T,T_z,r;i \right\vert
            H_{{\rm a}N}
              \left\vert T,T_z,r^\prime;j \right>_{N+1} \ \ \ .
\end{eqnarray}
with $X= g^\dagger_{N \sigma} g_{N+1 \sigma}
             +   g^\dagger_{N+1 \sigma} g_{N\sigma} $.
The first part simply gives the eigenstates of the previous iteration
\begin{eqnarray}
    \phantom{>}_{N+1} \left< T,T_z,r;i \right\vert H_N
     \left\vert T,T_z,r^\prime;j \right>_{N+1}
    = \delta_{rr^\prime}\delta_{ij} \nonumber \\
   \times \left\{ \begin{array}{rcl}
         E_N(T-1/2,r) &:& i=1 \\
         E_N(T-1/2,r) &:& i=2 \\
         E_N(T+1/2,r) &:& i=3 \\
         E_N(T+1/2,r) &:& i=4
          \end{array}
\right. \ \ .
\end{eqnarray}
To make use of the Wigner-Eckart Theorem 
\begin{eqnarray}
         \left< T,T_z,r;i \right\vert V^k_q
     \left\vert T^\prime,T_z^\prime,r^\prime;j \right> &=&
       \frac{1}{\sqrt{2T^\prime+1}}
        \big< T^\prime,r^\prime \big\vert \! \big\vert V_q^k
          \big\vert \! \big\vert T, r \big> \nonumber \\
           && \!\!\!\!\!\!\!\!\!\!\!\!
      \times   \big<T,T_z,k,q\vert T^\prime,T_z^\prime\big>
\end{eqnarray}
for the calculation of the second and third part, we have to define irreducible tensor
operators according to
\begin{eqnarray}
    \left[ T^\pm , V^k_q \right]_- &=& \sqrt{k(k+1) -q(q\pm1)}\ V^k_{q\pm1} \
\ ,\nonumber \\
    \left[ T_z , V^k_q \right]_- &=& q V^k_{q} \ \ \ .
\end{eqnarray}
In the standard case (where we have the spin $S$ instead of $T$ in eq.\ (29))
the tensor operators are just the annihilation operators $f_\downarrow$
and $f_\uparrow$ .
Here we have
\begin{eqnarray}
   V^0_{n,0} &=& \frac{1}{\sqrt{2}} \left(
                  (-1)^n g_{n\downarrow}^\dagger + g_{n\downarrow} \right) \ ,\nonumber \\
   V^1_{n,1} &=&  (-1)^{n+1} g_{n\uparrow}^\dagger \ \ \ ,\nonumber \\
   V^1_{n,0} &=& \frac{1}{\sqrt{2}} \left(
              (-1)^{n+1} g_{n\downarrow}^\dagger + g_{n\downarrow} \right) \ ,\nonumber \\
   V^1_{n,-1} &=&  - g_{n\uparrow} \ \ \ .
\end{eqnarray}
The  $g_{n\sigma}^\dagger$ and $g_{n\sigma}$ can be written in a form which
only depends on the operators $V^1_{n,0}$ and $V^0_{n,0}$
\begin{eqnarray}
   g_{n\uparrow}^\dagger &=& \frac{1}{\sqrt{2}}
                   (-1)^{n+1} \left[ T^+_n , V^1_{n,0} - V^0_{n,0} \right]_- \ \ \
\ \ ,\nonumber \\
   g_{n\uparrow} &=& \frac{-1}{\sqrt{2}}
                   \left[ T^-_n , V^1_{n,0} - V^0_{n,0} \right]_- \ \ \ ,\nonumber \\
   g_{n\downarrow}^\dagger &=& \frac{1}{\sqrt{2}}
                   (-1)^{n+1} \left( V^1_{n,0} - V^0_{n,0} \right) \ \ ,\nonumber \\
   g_{n\downarrow} &=& \frac{1}{\sqrt{2}}
                    \left( V^1_{n,0} + V^0_{n,0} \right) \ \ .
\end{eqnarray}
The matrix elements of eq.\ (26) are given in the appendix.

We now have all the information to set up the matrix $H_T(ri,r^\prime j)$.
The following steps are completely analogous to the standard case.
By diagonalization of $H_T(ri,r^\prime j)$ the Schr\"odinger equation of
the $(N+1)$st step is solved. Only a limited number of the lowest lying
states (500 in our calculations) are kept and used to set up the
Hamiltonian matrix for the next step.

\section{Discussion of the Fixed Points}

The various possible fixed points of the O(3)-AM can be identified by investigating
the flow diagrams for the lowest lying eigenenergies.
Fig.\ 1 shows a typical flow diagram for $V_{\rm a}=0$ 
(which corresponds to the standard SIAM).
The solid (dashed) lines belong to states with $T=0.5$
($T=1.5$). The ground state has $T=0.5$.
This flow diagram is similar to Fig.\ 6 in
\cite{Kri80} with the difference that in \cite{Kri80} the energies are labeled by
particle number $Q$ and spin $S$.
The system flows from the Free Orbital to the Strong Coupling fixed point.
The influence of the Local Moment fixed point is small for the
parameters used in Fig.\ 1 but increases for larger values of $U$.

In Fig.\ 2 we present a flow diagram for the isotropic case
$V_{\rm a} = V/2$.
We clearly find a new fixed point, which we call the non-Fermi liquid
fixed point (see the discussion in the following sections).
The non-Fermi liquid fixed point is unstable for small deviations from isotropy
as can be seen in Fig.\ 3 for $V_{\rm a} = 0.498 V$.
Up to iteration number 30 the flow diagrams of
Fig.\ 2 and 3 are almost identical
but after a crossover regime (iterations 30 to 50)
the system approaches the Fermi liquid fixed point of the
single channel case (see Fig.\ 1).
The interpretation of this as a Fermi liquid fixed point 
can also be found in Fabrizio et al.\ \cite{Fab95},
who investigated the two channel model by a bosonization method.

The static properties are calculated at each iteration step $N$ for the
temperature
\begin{equation}
    T_N = \frac{1}{\bar{\beta}}\Lambda^{-(N-1)/2} \ \ \ . \label{eq:T}
\end{equation}
For $\bar{\beta}$ we take the value
$\bar{\beta} = 0.46 $ (see the discussion in \cite{Kri80}).
Due to this connection of the number of iterations and the temperature
we can see that for $V_{\rm a} = 0.498 V$ there exists a high temperature regime
in which the system behaves similarly to the isotropic case and a low
temperature regime with a Fermi liquid ground state. This does not
automatically mean that we have non-Fermi liquid properties in the high
temperature regime because the non-Fermi liquid physics
itself dominates only in the very low temperature regime
of the isotropic case.

\section{Results for Entropy and Specific Heat}

At each step of the iterative diagonalization the impurity contribution
to the Free Energy is calculated
by
\begin{equation}
   F(T_N) = -k_{\rm B} T_N \left[ \ln \sum_i
        e^{-\bar{\beta}E_i} - \ln \sum_i e^{-\bar{\beta}E_i^0} \right] \ ,
\end{equation}
with $T_N$ given by equation (\ref{eq:T}).
The $E_i$ and $E_i^0$ are the scaled energies of the full system and the system
without impurity, respectively. The entropy $S(T)$ and the specific heat coefficient
$\gamma(T) = C(T)/T$ are calculated by numerically differentiating the
Free Energy
\begin{eqnarray}
 S(T) &=& -\frac{\partial}{\partial T} F(T)  \ ,\\
 \gamma(T) &=& -\frac{\partial^2}{\partial T^2} F(T) \ . 
\end{eqnarray}
It is useful to take the values of $F(T_N)$ only at odd (or even) $N$.
One has to be careful about the groundstate energy $E_{{\rm g},N}$ which is
substracted at each iteration step and has to be added again to give the 
correct derivative.
In the formula for the entropy
\begin{equation}
  S\left[ {\scriptstyle\frac{1}{2}}(T_N + T_{N-2})\right]
         = -\frac{F(T_N) + \Delta E  -F(T_{N-2})}
                      {T_N - T_{N-2}} \ \ ,
\end{equation}
the correction term $\Delta E$ takes the form
\begin{equation} 
         \Delta E =  \Lambda^{-(N-1)/2} \left( \sqrt{\Lambda} E_{{\rm g},N-1} +
                     E_{{\rm g},N} \right)\ \ .
\end{equation}
It consists of two terms because we compare the free energy
between {\it two} steps of the iteration procedure.

For the specific heat coefficient we use
\begin{eqnarray}
  \gamma\left[{\scriptstyle \frac{1}{2}}T_{N-2} +
              {\scriptstyle \frac{1}{4}}(T_N + T_{N-4})\right] =\nonumber \\
       \frac{S\left[\frac{1}{2}(T_N + T_{N-2})\right]-
               S\left[\frac{1}{2}(T_{N-2} + T_{N-4})\right] }{T_N - T_{N-4}} \ .
\end{eqnarray}
Results for the entropy are shown in Fig.\ 4. For $V_{\rm a}=0$ 
the entropy vanishes for $T\to 0$. In the isotropic case $V_{\rm a} =V/2$
we find a zero point entropy of 
$1/2 \ln 2$ (within the numerical accuracy).
For any $0<V_{\rm a} < V/2 $  we find the behaviour that has already been discussed
in the previous section. The entropy is identical to the result of
the isotropic case in a high temperature regime, but
it tends to zero for $T\to 0$.

The specific heat coefficient $\gamma(T)$ is plotted in Fig.\ 5 for
$V_{\rm a}=0$, $V_{\rm a}=V/2$ and $V_{\rm a}=0.43 V$.
In the single band limit we find the usual Fermi liquid behaviour with
a constant $\gamma$ at low temperatures, while in the isotropic case
there is a very clear logarithmic dependence.
For any $V_{\rm a} \ne V/2$, $\gamma(T)$ approaches  a constant value ($\bar{\gamma}$)
for $T\to 0$. The energy scale $T^*$ (defined as $1/\bar{\gamma}$)
depends quadratically on $(V-2V_{\rm a})$ (not plotted here).
This result corresponds to the calculations by Pang and Cox \cite{Pan91}
who found a $T^* \propto (\Delta J)^2$ dependence for the two channel Kondo
model with channel anisotropy.

In the isotropic case we define the energy scale $T_{\rm i}= 1/c$
with $c$ the prefactor of the $\ln T$ term in $\gamma$. In the large
$U$ limit we find $T_{\rm i} \propto \exp( -{\rm const}\cdot U/\pi \Delta)$
(see Fig.\ 6, $\Delta = 1/2 \pi V^2$) which corresponds to the expression for the Kondo temperature in the 
two channel Kondo model ($T_{\rm K} 
 \propto \exp( {\rm -const}/J)$, $J\propto \Delta/U$).
For $U=0$ the prefactor $c$ vanishes and $\gamma$ approaches a
constant value proportional to $\bar{T}^{-1}$, with $\bar{T}$
a second energy scale not present in the  large
$U$ limit. In the
small $U$-regime the energy scale $T_{\rm i}$ diverges as $U^{-2}$ 
(see Fig.\ 6). This result is in agreement with second order perturbation
theory around  $U=0$ \cite{Col95a,Zha96}.

\section{Results for `Spin plus Isospin' Susceptibility and Wilson ratio}

The most relevant susceptibility to calculate is $\chi^\prime=
\chi_{\sigma,{\rm imp}} + \chi_{\sigma,{\rm c}} +\chi_{\tau,{\rm c}}$,
which includes spin and isospin of the conduction electrons but only
the spin of the impurity. $\chi^\prime$ is related to the spin 
susceptibility of the two channel Kondo model where 
$\chi_{\sigma,{\rm c}}$ and  $\chi_{\tau,{\rm c}}$ represent the spin
susceptibility of channel one and two, respectively.
The quantity $\chi^\prime$, however, is not conserved and therefore
difficult to calculate within the NRG (additional matrix elements have to be
calculated iteratively).

On the other hand, the `spin plus isospin' susceptibility 
$\chi_T=
\chi_{\sigma,{\rm imp}} + \chi_{\tau,{\rm imp}} + 
\chi_{\sigma,{\rm c}} +\chi_{\tau,{\rm c}}$ 
is conserved and hence can be calculated from the energy states only
using the relation
\begin{equation}
\chi_T = \beta \left< T_z^2\right> \ ,
\end{equation}
\begin{equation}
\left< T_z^2\right> = \frac{\sum_i T_z^2 \exp(-\bar{\beta}E_i)}
                   {\sum_i \exp(-\bar{\beta}E_i)} -
                      \frac{\sum_i T_z^2 \exp(-\bar{\beta}E_i^0)}
                   {\sum_i \exp(-\bar{\beta}E_i^0)} \  \ \ ,
\end{equation}
with $T_z$ the $z$-component of the total `spin plus isospin' $T$.

In the limit of large $U$ the susceptibility $\chi_{\tau,{\rm imp}}$
vanishes because charge fluctuations (described by
the isospin $\tau$) are frozen out at the impurity site.
That means that, although we are studying the susceptibility $\chi_T$
for all values of $U$, it corresponds to  $\chi^\prime$ in the
limit where O(3)-AM and ($\tau$-$\sigma$)-model are related via
the Schrieffer-Wolff transformation.

In the isotropic case ($V_{\rm a} = V/2$) we find $\chi_T\propto \ln T$
while for any $V_{\rm a} \ne V/2$, $\chi_T$ approaches a constant value
for $T\to 0$ (not plotted here).

The Wilson ratio  $\chi_T /\gamma$ is plotted in Fig.\ 7 for both 
$V_{\rm a}=0$ and $V_{\rm a}=V/2$. Also shown is the result for $\chi_S /\gamma$
($\chi_S = \chi_{\sigma,{\rm imp}} + \chi_{\sigma,{\rm c}}$)
calculated with the standard program for the SIAM. In contrast to $\chi_S /\gamma$
the `spin plus isospin' Wilson ratio is independent of $U$ (within the numerical
accuracy). For $V_{\rm a}=V/2$ we find a Wilson ratio which
is consistent with the value of 8/3 of the
two channel Kondo model, within the numerical accuracy.

For $0 \le V_{\rm a}<V/2$ the Wilson ratio is a nonuniversal function
of $V_{\rm a}$ (see Fig.\ 8).
Due to the vanishing energy scale in this limit 
the numerical calculation of $\chi_T$ and $\gamma$ 
gets more and more inaccurate. However, it is clear that $R$
decreases rapidly as $V_{\rm a}\to V/2$ (note that the 
$\chi_T/\gamma$-ratio is discontinuous at this point and for 
$V_{\rm a}= V/2$ the Wilson ratio takes the value $8/3$).

\section{Summary}

To summarize, we have discussed a numerical renormalization group
approach to the O(3)-symmetric Anderson model that can be related
to the ($\tau$-$\sigma$)-model introduced by Coleman et al.\
\cite{Col95} via a Schrieffer-Wolff transformation.
Both spin and charge are not conserved in the O(3)-symmetric model
so we have to use the only conserved quantity `spin plus isospin'
to classify the eigenstates in the iterative diagonalization procedure.

The flow diagrams show a non-Fermi liquid fixed point
for $V_{\rm a} = V/2$ (corresponding to $J_1 = J_2$ in the
($\tau$-$\sigma$)-model)
which is unstable for any small deviations from the isotropic
case. 
All static properties calculated
for $V_{\rm a} = V/2$ show the non-Fermi liquid behaviour of the 
two channel Kondo model:
\begin{itemize}
\item a zero point entropy of $\frac{1}{2}\ln 2$,
\item a logarithmic temperature dependence of the specific heat
       $C(T) \propto T\ln T$ ,
\item a logarithmic temperature dependence of the `spin plus
      isospin' susceptibility $\chi_T (T) \propto \ln T $ ,
\item a Wilson ratio of $\chi_T/\gamma=8/3$ independent of $U$.
\end{itemize}
The relation of these results to those of the Conformal Field
Theory and Perturbation Theory and the interpretation of the 
non-Fermi liquid fixed point in terms of Majorana Fermions will
be dealt with in a separate publication.

For $V_{\rm a} \ne V/2$ the model shows the conventional Fermi liquid
behaviour similar to the standard SIAM. There is no indication of a new non-Fermi liquid fixed
point as conjectured in \cite{Col95a}. We find an energy scale
$T^* \propto (J_1 -J_2)^2$ which corresponds to the result of
Pang and Cox \cite{Pan91}. This energy scale should correspond to 
$T_{\rm a}$, the anisotropic energy scale in \cite{And95},
but there $T_{\rm a} \propto (J_1 -J_2)$. The difference is probably due to the 
different cut-off scheme used (giving a different prefactor to the exponential).

It is interesting to see whether the behaviour of non-conserved quantities
(such as charge- and spin-susceptibility), dynamic and transport
properties are also the same for the O(3)-AM and the two channel Kondo model.
The generalization of the NRG for the calculation of these quantities is in
progress.

We wish to thank Th.\ Pruschke, G.\ M.\ Zhang and J.\ Keller
for a number of stimulating discussions. One of us (R.B.) was supported
by a grant from the Deutsche Forschungsgemeinschaft, grant No.\ Bu965-1/1 
and we also acknowledge support from the EPSRC (grant No.\ GR/J85349).

\onecolumn

\appendix
\section{}

In this appendix we want to list the results for the matrix elements
necessary to set up the Hamiltonian matrix (25).
For the matrix element
\begin{equation}
\sum_\sigma
    \phantom{>}_{N+1} \left<T,T_z,r;i \right\vert \Big(
                g^\dagger_{N \sigma} g_{N+1 \sigma}
             +   g^\dagger_{N+1 \sigma} g_{N\sigma}  \Big)
              \left\vert T,T_z,r^\prime;j \right>_{N+1}
\end{equation}
we get the following results

$i=2,\ j=1$:
\begin{equation}
  \frac{1-2T}{4\sqrt{T}\sqrt{T^2-1/4}}
      { }  _N\big< T -1/2,r \big\vert \! \big\vert V_{N,0}^1
          \big\vert \! \big\vert T-1/2, r^\prime \big>_N
  +  \frac{1}{2\sqrt{T}}
      { }  _N\big< T -1/2,r \big\vert \! \big\vert V_{N,0}^0
          \big\vert \! \big\vert T-1/2, r^\prime \big>_N ;
\end{equation}

$i=2,\ j=4$:
\begin{equation}
    (-1)^N \frac{1}{\sqrt{2T+1}}
      { }  _N\big< T -1/2,r \big\vert \! \big\vert V_{N,0}^1
          \big\vert \! \big\vert T+1/2, r^\prime \big>_N;
\end{equation}

$i=3,\ j=1$:
\begin{equation}
    -\frac{1}{\sqrt{2T+1}}
        { } _N\big< T +1/2,r \big\vert \! \big\vert V_{N,0}^1
          \big\vert \! \big\vert T-1/2, r^\prime \big>_N;
\end{equation}

$i=3,\ j=4$:
\begin{equation}
  (-1)^N
  \frac{\sqrt{2T+3}}{2\sqrt{T+1}\sqrt{2T+1}}
        { } _N\big< T +1/2,r \big\vert \! \big\vert V_{N,0}^1
          \big\vert \! \big\vert T+1/2, r^\prime \big>_N \nonumber
\end{equation}
\begin{equation}
  + (-1)^N \frac{1}{2\sqrt{T+1}}
        { } _N\big< T +1/2,r \big\vert \! \big\vert V_{N,0}^0
          \big\vert \! \big\vert T+1/2, r^\prime \big>_N \ .
\end{equation}

The remaining matrix elements ($i=1$, $j=2$ etc.) follow from
the hermiticity of $H(ri,r^\prime j)$.
The anomalous part in the Hamiltonian matrix
\begin{equation}
\phantom{>}_{N+1} \left<T,T_z,r;i \right\vert
            H_{{\rm a}N}
              \left\vert T,T_z,r^\prime;j \right>_{N+1}
\end{equation}
contains the contributions

$i=2,\ j=1$:
\begin{equation}
    \frac{1}{\sqrt{T}}
        { } _N\big< T -1/2,r \big\vert \! \big\vert V_{N,0}^0
          \big\vert \! \big\vert T-1/2, r^\prime \big>_N \ ;
\end{equation}

$i=3,\ j=4$:
\begin{equation}
   (-1)^N \frac{1}{\sqrt{T+1}}
        { } _N\big< T +1/2,r \big\vert \! \big\vert V_{N,0}^0
          \big\vert \! \big\vert T+1/2, r^\prime \big>_N \ .
\end{equation}
The next step is to relate the reduced matrix elements
$ _N<|\!|V_{N,0}^0|\!|>_N$ and  $ _N<|\!|V_{N,0}^1|\!|>_N$
to the unitary matrices $U_T$ of the previous iteration
(the $U_T$ diagonalize the Hamiltonian matrices $H_T$).
The results are

\begin{displaymath}
  _N\big< T ,w \big\vert \! \big\vert V_{N,0}^1
          \big\vert \! \big\vert T+1, w^\prime \big>_N
       =   \hspace{11cm}  
\end{displaymath}
\begin{equation}
          (-1)^{N+1} \frac{\sqrt{2T+1}\sqrt{2T+3}}{\sqrt{2T+2}} 
        \sum_r \left[ U_T (w,r4) U_{T+1} (w^\prime,r2) +
             U_T (w,r3) U_{T+1} (w^\prime,r1)
        \right] \ \ ,
\end{equation}

\begin{displaymath}
  _N\big< T ,w \big\vert \! \big\vert V_{N,0}^1
          \big\vert \! \big\vert T, w^\prime \big>_N
       =   \hspace{11cm}   
\end{displaymath}
\begin{displaymath}
       =  \frac{\sqrt{2T+1}\sqrt{T+1}}{\sqrt{2T}}
       \sum_r \left[ (-1)^N U_T (w,r2) U_T (w^\prime,r1) -
             U_T (w,r1) U_T (w^\prime,r2)
        \right]
\end{displaymath}
\begin{equation}
       +  \frac{\sqrt{2T+1}\sqrt{T}}{\sqrt{2T+2}}
       \sum_r \left[ U_T (w,r3) U_T (w^\prime,r4) +
             (-1)^{N+1} U_T (w,r4) U_T (w^\prime,r3)
        \right]  \ \  ,
\end{equation}

\begin{displaymath}
  _N\big< T ,w \big\vert \! \big\vert V_{N,0}^1
          \big\vert \! \big\vert T-1, w^\prime \big>_N
       =   \hspace{11cm}   
\end{displaymath}
\begin{equation}
       = - \frac{\sqrt{2T+1}\sqrt{2T-1}}{\sqrt{2T}}
       \sum_r \left[ U_T (w,r2) U_{T-1} (w^\prime,r4) +
             U_T (w,r1) U_{T-1} (w^\prime,r3)
        \right] \  ,
\end{equation}

\begin{displaymath}
  _N\big< T ,w \big\vert \! \big\vert V_{N,0}^0
          \big\vert \! \big\vert T, w^\prime \big>_N
       =  \hspace{11cm}   
\end{displaymath}
\begin{displaymath}
       = \sqrt{T+\frac{1}{2}}
       \sum_r \left[ (-1)^N U_T (w,r2) U_T (w^\prime,r1) -
             U_T (w,r3) U_T (w^\prime,r4)
        \right.
\end{displaymath}
\begin{equation}
        \left. + U_T (w,r1) U_T (w^\prime,r2) +
             (-1)^{N+1} U_T (w,r4) U_T (w^\prime,r3)
        \right]
\end{equation}
\twocolumn

\begin{figure}[htb]
\epsfxsize=3.4in
\epsffile{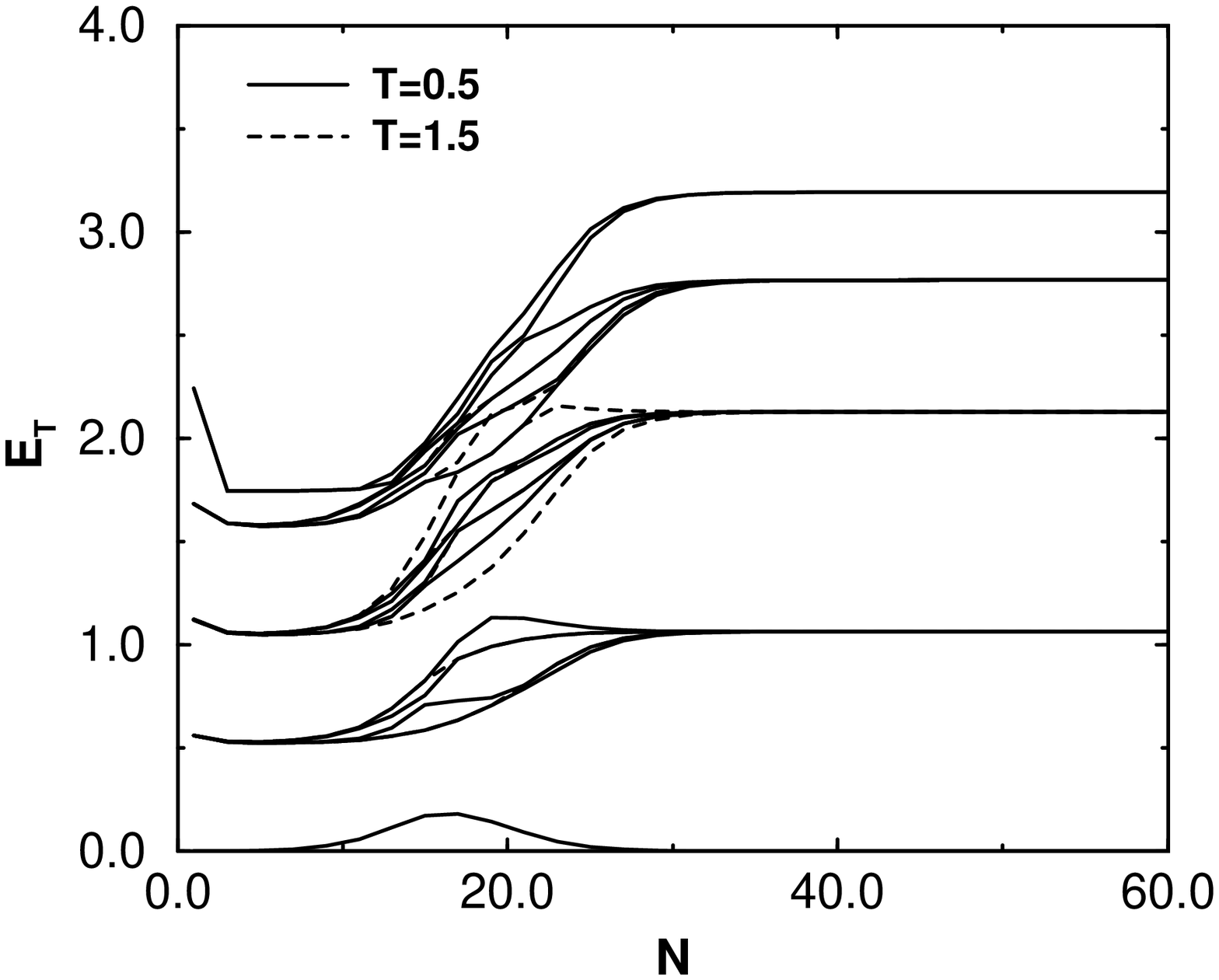}
\caption{Flow diagram for $\varepsilon_{\rm f} = -0.000714$, $U=-2\varepsilon_{\rm f}$,
               $V=0.01414$ and $V_{\rm a} = 0$. Solid and dashed lines belong to
               $T=0.5$ and $T=1.5$, respectively. The system flows
               to the Fermi liquid fixed point.}
\end{figure}

\begin{figure}[htb]
\epsfxsize=3.4in
\epsffile{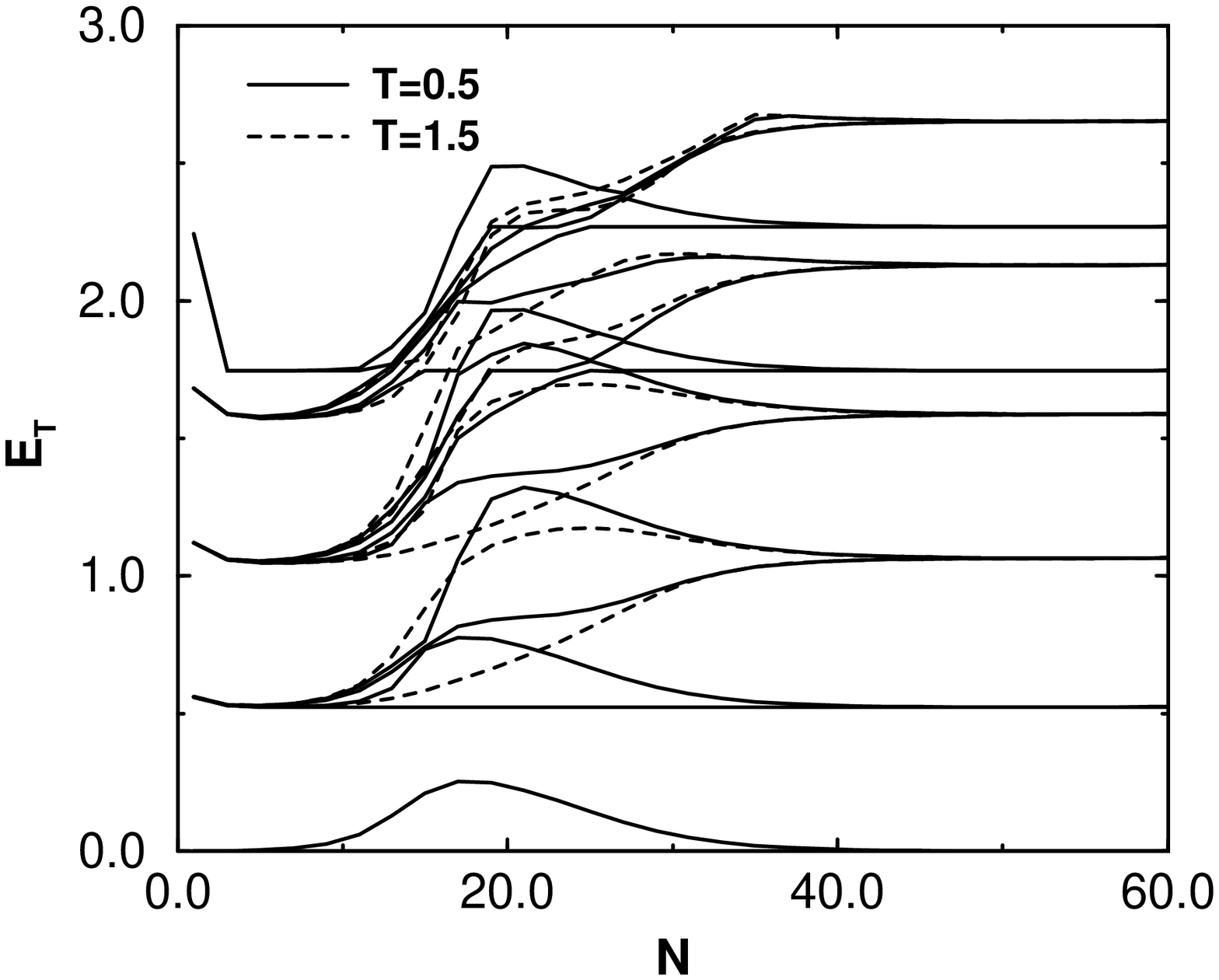}
\caption{Flow diagram for $\varepsilon_{\rm f} = -0.000714$, $U=-2\varepsilon_{\rm f}$,
               $V=0.01414$ and $V_{\rm a} = V/2$. Solid and dashed lines belong to
               $T=0.5$ and $T=1.5$, respectively.  The system flows
               to the non-Fermi liquid fixed point.}
\end{figure}

\begin{figure}[htb]
\epsfxsize=3.4in
\epsffile{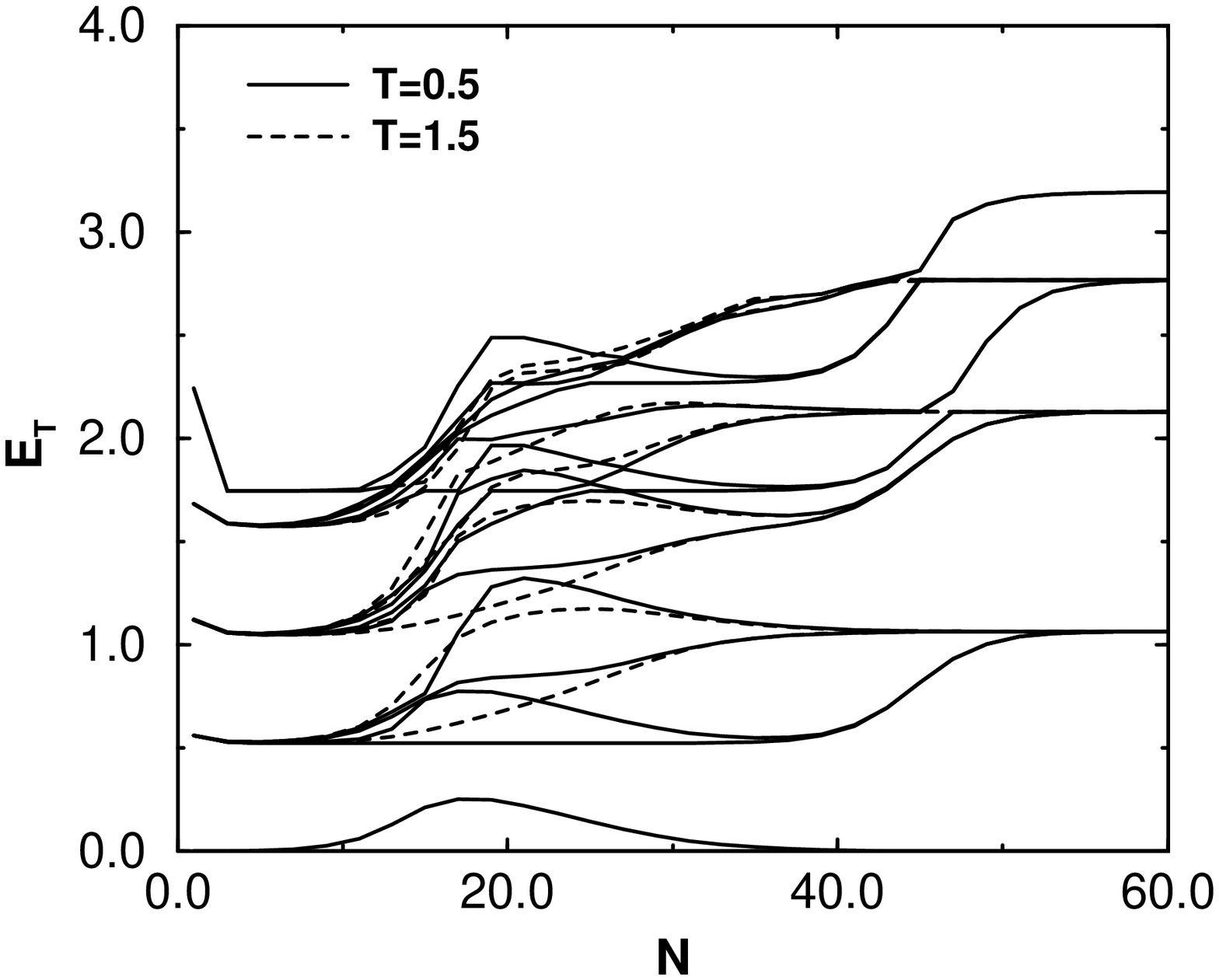}
\caption{Flow diagram for $\varepsilon_{\rm f} = -0.000714$, $U=-2\varepsilon_{\rm f}$,
               $V=0.01414$ and $V_{\rm a} = 0.00704$. Solid and dashed lines belong to
               $T=0.5$ and $T=1.5$, respectively.   After a crossover regime, the system flows
               to the Fermi liquid fixed point.}
\end{figure}

\begin{figure}[htb]
\epsfxsize=3.4in
\epsffile{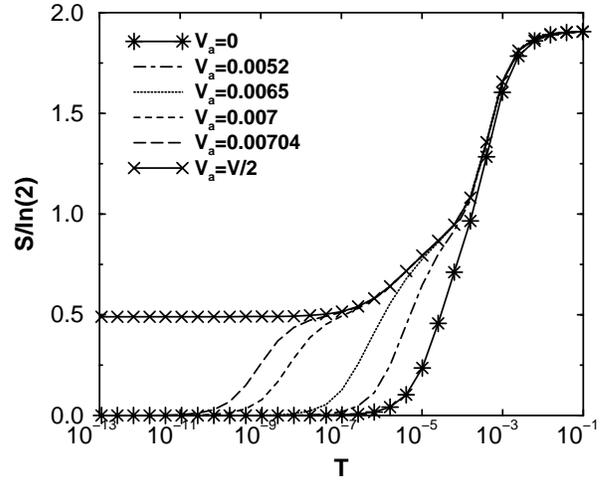}
\caption{Entropy for $\varepsilon_{\rm f} = -0.000714$, $U=-2\varepsilon_{\rm f}$,
               $V=0.01414$ and $V_{\rm a}=0$, $V_{\rm a}=V/2$, and various
               values of  $0< V_{\rm a} < V/2$ . In the isotropic case we find the
               residual entropy of $1/2 \ln 2$.}
\end{figure}

\begin{figure}[htb]
\epsfxsize=3.4in
\epsffile{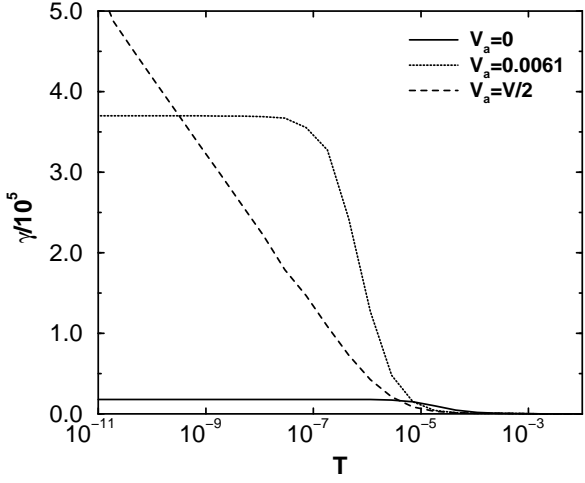}
\caption{Specific heat coefficient $\gamma(T)$ for
               $\varepsilon_{\rm f} = -0.000714$, $U=-2\varepsilon_{\rm f}$,
               $V=0.01414$ and $V_{\rm a}=0$ (solid line), $V_{\rm a}=0.0061$ (dotted line)
               and $V_{\rm a}=V/2$ (dashed line).}
\end{figure}

\begin{figure}[htb]
\epsfxsize=3.4in
\epsffile{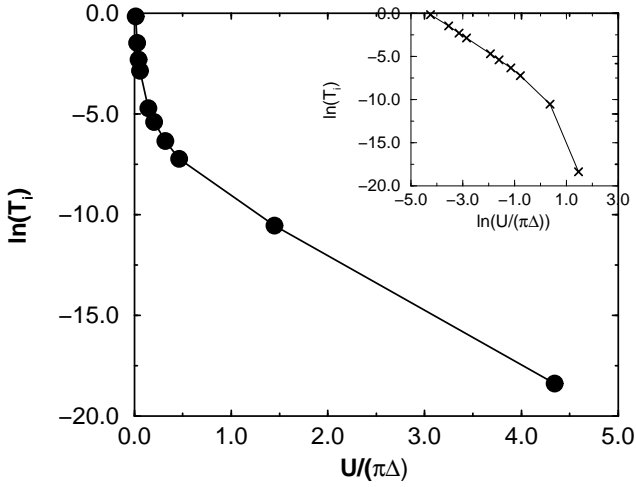}
\caption{Energy scale in the isotropic case $V_{\rm a} = V/2$. For small $U$,
           $T_{\rm i}$ is proportional to $U^{-2}$ (insert) but goes as 
          $\exp(-{\rm const} \cdot  U/\pi\Delta)$ for large $U$.}
\end{figure}

\begin{figure}[htb]
\epsfxsize=3.4in
\epsffile{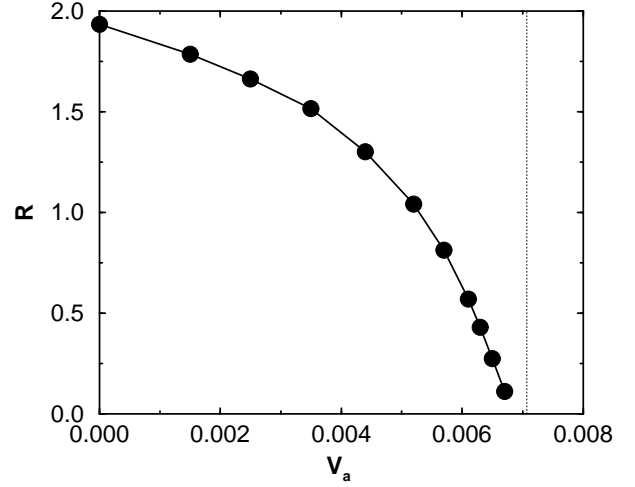}
\caption{$V_{\rm a}$-dependence of the Wilson Ratio $R$ for 
         $\varepsilon_{\rm f} = -0.000714$, $U=-2\varepsilon_{\rm f}$
          and $V=0.01414$. 
        The $V_{\rm a}\!\to\!
         0$-case corresponds to the  ordinary SIAM. The numerical calculation
         of $R$ becomes more and more inaccurate as $V_{\rm a}$ approaches $V/2$,
         (vertical line) but the data indicate that $R$ vanishes in this limit.}
\end{figure}

\begin{figure}[htb]
\epsfxsize=3.4in
\epsffile{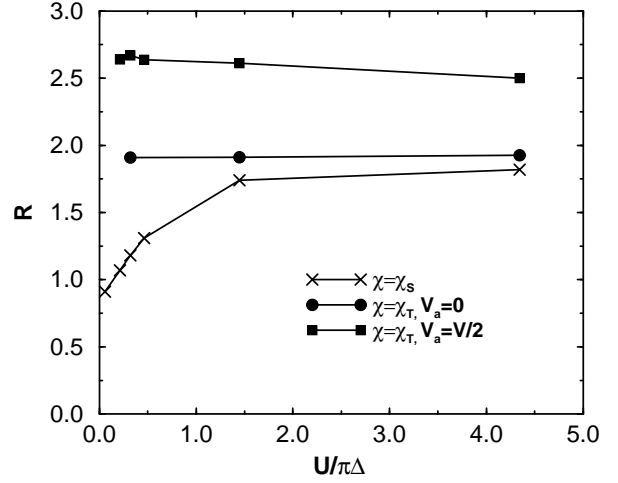}
\caption{Wilson Ratio $\chi_T /\gamma$ for $V_{\rm a}=0$ 
               and $V_{\rm a}=V/2$ . Also shown is the Wilson Ratio 
               $\chi_S /\gamma$ for the standard SIAM.}
\end{figure}

\end{document}